%% file: paper.tex
\documentclass[conference]{IEEEtran}

\usepackage{graphicx}
\usepackage{amsmath}
\usepackage{amssymb}
\usepackage{algorithm}
\usepackage{algorithmicx}
\usepackage{algpseudocode}
\usepackage{amsfonts,amssymb,amsthm}
\usepackage{xspace}
\usepackage[acronym,nomain]{glossaries}
\usepackage{caption}
\usepackage{listings}
\usepackage{blindtext}
\usepackage{svn-multi}
\usepackage{multirow}
\usepackage{lipsum}
\usepackage{tikz}
\usepackage{url}
\usetikzlibrary{patterns}
\usepackage{subcaption}
\newcount\outlineon
\newcommand{\outline}[1]{\ifnum\outlineon>0\
	\\\noindent\fbox{\begin{minipage}{\linewidth}{\footnotesize#1}\end{minipage}}\\\\\fi}
\newcommand{\execlude}[1]{}

\newcommand{\proofend}[2]{ $ \blacksquare$ \vspace{5mm} }

\theoremstyle{plain}

\theoremstyle{definition}

\theoremstyle{remark}

\newif\ifcomments

\commentstrue 
\commentsfalse 

\input{comments}

\makeglossaries
\IEEEoverridecommandlockouts
\begin{document}
	\ifcomments
	\outlineon = 1
	\fi
	
	%
\title{Energy-aware Time- and Event-triggered \\KVM Nodes}


\author{\IEEEauthorblockN{Isser Kadusale, Gautam Gala and Gerhard Fohler}
	\IEEEauthorblockA{Chair of Real-time Systems,\\
		University of Kaiserlautern-Landau (RPTU), Germany\\
		\{kadusale,gala,fohler\}@eit.uni-kl.de\\
	}
}		
\maketitle

\ggcomment{Overall 8 pages}
\input{abstract}


\vspace{-0.25cm}
\input{Section_Introduction.tex}

\input{Section_Related_Work.tex}

\input{Section_System_Model.tex}

\input{Section_Scheduler.tex}

\input{Section_Linux_Implementation.tex}


\input{Section_Evaluation.tex}


\input{Section_Conclusion.tex}

\bibliographystyle{IEEEtran}      
\bibliography{references}   

\end{document}

%% file: comments.tex
\ifcomments
	\usepackage[colorinlistoftodos]{todonotes}
\else
	\usepackage[disable]{todonotes}
\fi

\newcommand{\ggcomment}[1]{\todo[inline, color=green!40, author=GG]{#1}}



%% file: abstract.tex
\begin{abstract}
Industries are considering the adoption of cloud and edge computing for real-time applications due to current improvements in network latencies and the advent of Fog and Edge computing.
Current cloud paradigms are not designed for real-time applications, as they neither provide low latencies/jitter nor the guarantees and determinism required by real-time applications.
Experts estimate that data centers use 1\% of global electricity for powering the equipment, and in turn, for dealing with the produced heat.
Hence, energy consumption is a crucial metric in cloud technologies.
Applying energy conservation techniques is not straightforward due to the increased scheduling overheads and application execution times.
Inspired by slot shifting, we propose an algorithm to support energy-aware time-triggered execution of periodic real-time VMs while still providing the ability to execute aperiodic real-time and best-effort VMs in the slack of the time-triggered ones.
The algorithm considers energy reduction techniques based on dynamic power management and dynamic voltage and frequency scaling.
We implement our algorithm as an extension to the Linux kernel scheduler (for use with the KVM hypervisor)
and evaluate it on a server-grade Intel Xeon node.

\end{abstract}

%% file: Section_Introduction.tex
\section{Introduction}
\label{section:introduction}

The usage of cloud technologies is continuously rising. The cloud offers better hardware efficiency and maintainability, higher availability, and more cost-effective scaling compared to traditional systems. For these reasons, cloud computing is picking up steam in safety-critical domains such as in railway \cite{secredas, siemensinfrastructureinthecloud, SIL4Cloud, thales2022research}, automotive \cite{soafee.io}, and smart manufacturing \cite{eker_angelsmark_sefidcon_2020}. However, safety-critical applications in these domains require timing guarantees that current cloud offerings cannot fulfill.

Time-triggered scheduling is widely used in the safety-critical domain because of its determinism and strong isolation properties. It can be used in cloud environments to help fulfill the timing demands of safety-critical applications. Specialized hypervisors, such as XtratuM \cite{xtratum}, support time-triggered VMs but are proprietary and unsuitable for cloud computing. Previous work for time-triggered generic hypervisors, such as Tableau \cite{DBLP:conf/eurosys/VangaGB18} and ARINC-653 scheduler for Xen \cite{studer2019}, support time-triggered VMs in a private cloud. However, these approaches do not allow adding VMs to the existing scheduling table or have a high overhead table regeneration process. Moreover, previous work demonstrated that the KVM (with PREEMPT\_RT) hypervisor \cite {kvm}, in general, has considerably lower latency as compared to Xen, making KVM better suited to time-sensitive and safety-critical applications \cite{gala2021rt, Abeni2020kvm_xen, Deshane2008QuantitativeCO}. In addition, most time-triggered solutions undermine the benefits of cloud computing since they do not utilize the unused resources from time-triggered VMs.

Another aspect that existing time-triggered cloud hypervisors overlook is energy. High energy usage negatively impacts the environment, and energy consumption accounts for a significant portion of the costs of a data center \cite{greenberg2008cost}. Ignoring energy in computing systems also leads to more heat generation, increasing cooling requirements, and decreasing reliability and longevity. Therefore, it is essential to use energy-reduction techniques in computing systems. However, applying energy-reduction techniques to the real-time cloud adds another dimension to the problem, since they impact scheduling overheads and application execution times. These techniques must maintain timing guarantees, especially for safety-critical applications.

The slot shifting algorithm \cite{fohler1995joint} combines time-triggered and event-triggered scheduling to get the benefits of both. It allows flexibility in a time-triggered schedule while maintaining its determinism and isolation properties. Slot shifting can accept and guarantee execution time to event-triggered tasks by keeping track of the available slack in the time-triggered schedule. This system of tracking the available slack could also be extended and used to ensure that energy-reduction techniques do not cause timing violations.

Inspired by the slot-shifting algorithm, we propose a novel energy-aware scheduling algorithm for cloud nodes to ensure that safety-critical VMs meet their requirements. Utilizing the concepts in slot shifting allows for having a time-triggered schedule with the flexibility of handling event-triggered tasks. In summary, our scheduler:
\vspace{-0.2cm}
\begin{itemize}
	\item supports time-triggered and event-triggered safety-critical/time-sensitive VMs.
	\item uses energy-reduction techniques to conserve energy without impacting the timing guarantees of safety-critical VMs.
	\item exploits the slack of the time-triggered VMs to execute best-effort/non-critical VMs.
	\item has low overheads when adding new VMs by appending them to the existing scheduling table at run-time instead of completely regenerating it.
	\item can be plugged into an RT-cloud resource orchestration solution such as those proposed in \cite{gala2021rt, gala2021work}.
\end{itemize}
Finally, we implemented our scheduler inside the Linux Kernel. Our approach keeps run-time overhead low by separating the TT task dispatching via a newly created scheduling class (on TT/managed cores) from the energy-aware slot-shifting algorithm decision-making. We execute the energy-aware slot-shifting algorithm on non-TT/non-managed cores with a period equal to the slot length. We experimentally evaluated server-grade Dell hardware with an Intel 2$^{nd}$ generation Xeon processor (Cascade Lake).

The remainder of the paper is as follows: Section \ref{section:related_work} presents related work. Section \ref{section:system_model} introduces the used system models. Section \ref{section:scheduler} and Section \ref{section:linux_implementation} describe the proposed scheduler and its implementation for the Linux kernel. Section \ref{section:evaluation} presents our evaluation, and \ref{section:conclusion} concludes the paper.

%% file: Section_Related_Work.tex
\section{Related Work}
\label{section:related_work}
\paragraph{Joint scheduling of TT and ET RT tasks}
Many existing works have explored joint time-and event-triggered approaches, e.g.,~\cite{fohler1995joint, schorr2015adaptive, real}.
However, they do not integrate these approaches to combine TT and ET tasks with the Linux kernel scheduler. Fohler~\cite{fohler1995joint} presented the Slot-shifting algorithm for joint scheduling of TT and ET tasks. Schorr~\cite{schorr2015adaptive} presented a multiprocessor extension to the slot-shifting algorithm. We further extend the slot-shifting algorithm and propose two approaches to integrate energy awareness (using DPM and DVFS).
\paragraph{Dynamic Voltage and Frequency Scaling (DVFS) techniques}
Early research on energy-aware real-time scheduling explored the use of Dynamic Voltage and Frequency Scaling (DVFS) techniques which adjust the voltage and frequency of the processor to save energy \cite{DBLP:conf/focs/YaoDS95}. DVFS involves identifying the slack in a schedule and using that information to allow the execution of tasks with a lower processor frequency without violating their deadlines. Previous works (e.g., \cite{DBLP:conf/ecrts/AydinMMM01}) proposed offline algorithms determining the optimal execution frequency for tasks in a task set, assuming that the tasks execute until their WCET. Dynamic approaches, such as \cite{DBLP:conf/rtss/LeeS04, DBLP:conf/iceac/BambaginiPMB11}, take advantage of the unused execution time of tasks finishing earlier than their WCETs.

\paragraph{Dynamic Power Management (DPM) techniques}
The power consumption of a processor has a dynamic and static component. The dynamic power consumption is due to circuit switching activity, while the static power consumption is due to leakage current. 

Unlike DVFS, Dynamic Power Management (DPM) techniques reduce static power consumption by switching the processor to a lower power mode. However, while in a low-power mode, the processor cannot execute any tasks, and transitioning to and from a low-power mode incurs a time and energy overhead. Processors typically have several low-power modes, each with a different time and energy overhead. In general, modes with lower power consumption have higher overheads. Energy-aware real-time scheduling algorithms \cite{DBLP:conf/ecrts/LeeRK03, DBLP:conf/ecrts/AwanP11} consolidate idle periods so that the processor can be put in a lower power mode and minimize transition overheads.

\paragraph{Energy-saving real-time approaches for clouds}

Chen et al. \cite{DBLP:conf/hpcc/ChenZZW13} proposed an energy-aware scheduling scheme for scheduling jobs in a virtualized cloud. The proposed method dynamically creates, deletes, and consolidates VMs to meet real-time requirements while saving energy. Zhu et al. \cite{DBLP:journals/tcc/ZhuYCWYL14} presented an energy-aware scheduling architecture for real-time tasks in virtualized clouds. They employ a rolling horizon for waiting tasks, allowing rescheduling for better system schedulability and less energy consumption. These works, however, only handle aperiodic tasks. Sun and Cho \cite{sun2022lightweight} proposed a real-time scheduling algorithm for cloud computing servers. They use flow networks to schedule periodic tasks, ensuring the system meets the deadlines while consolidating idle periods to maximize energy savings using DPM. The proposed scheduler does not consider aperiodic tasks. In addition to only being limited to a single type of task, these schedulers cannot offer the same determinism that a time-triggered scheduler offers.

\paragraph{Energy-saving techniques in Linux/KVM}
Linux/KVM offers dynamic CPU frequency scaling (DVFS) via the CPUFreq subsystem. Linux frequency scaling governors implement the algorithms to compute the desired CPU frequency. At the same time, the scaling drivers interact with the CPU directly to enact the frequencies computed by the current governor.

Linux/KVM offers dynamic power management (DPM) that turns off the power or switches system components to a low-power state when inactive/idle for a pre-defined minimum amount of time.

Unfortunately, these energy-saving features in the Linux kernel need to be often disabled for use with real-time systems as they impact the kernel latency and determinism.

\paragraph{Energy-aware Time-triggered hypervisors/OSes}
Proprietary real-time hypervisors, such as XtratuM \cite{xtratum}, often support energy-aware (DPM) table-driven scheduling but are not suitable for use in cloud environments as they have limited hardware and existing software support, allow only a handful of guest operating systems and have other limitations such as limited CPU models for VMs.

Xen with ARINC 653 scheduler \cite{studer2019} and Tableau \cite{DBLP:conf/eurosys/VangaGB18} extension to Xen hypervisor introduce support for table-driven scheduling of VMs but are not energy-aware.
Moreover, authors in \cite{gala2021rt} demonstrated that Xen, in general, has higher overheads than Linux (+ PREEMPT\_RT patch) in conjunction with KVM, and therefore, KVM is better suited for low latency real-time safety-critical applications.

Very recently, Karachatzis et al.~\cite{schedTTTech} presented an implementation and evaluation of a kernel-level time-triggered scheduling approach for Linux processes. However, the approach turns off energy-saving features in the Linux kernel to reduce the kernel's latency and improve determinism. In addition to energy awareness in Linux/KVM, our approach supports the guaranteed execution of ET RT VMs by appending them to the scheduling table at run-time (without needing to regenerate the table).

Gala et al. \cite{gala2023joint} proposed a new TT scheduling class for the Linux kernel and integrated slot-shifting to support guaranteed execution of ET RT tasks. We build upon this work to support TT execution VMs (in conjunction with KVM). Additionally, we integrate our new slot-shifting extension to support energy awareness while providing flexibility.

%% file: Section_System_Model.tex
\section{Models}
\label{section:system_model}

We consider a time-triggered approach and assume a timeline that is divided into equal lengths called slots. A slot is a time interval, $[t, t+1)$, where $t$ is the slot length. Slots are synchronized between all cores, and scheduling decisions take effect at the start of each slot.

We adopt the same model for VMs and their mapping to periodic tasks as in Tableau \cite{DBLP:conf/eurosys/VangaGB18} as well as their mapping to periodic real-time tasks. Each VM consists of one or more vCPUs and each vCPU has a reserved utilization, $U$, and maximum scheduling latency, $L$. Each vCPU is then mapped to a periodic task, $\tau_i$, with a corresponding worst-case execution time (WCET), $C_i$, and period, $T_i$, and where the relative deadline is equal to the period. Since a task can be scheduled immediately at the start or right at the end of a period, the longest gap between two invocations of a task is $2(T-C)$, or equivalently $2T(1-U)$. This would be the maximum scheduling latency experienced by the vCPU. Therefore, in order to meet the latency requirement, a suitable period, $T$, must be selected that satisfies Equation \ref{eq:T}.
\begin{align}
	T \le \frac{L}{2(1-U)} \label{eq:T}
\end{align}
With the period, $T$, selected, the WCET can be calculated as $C=UT$ to meet the required utilization. We assume that the specified utilization, $U$, and by extension the resulting WCET, $C$, takes into account possible interference from other cores.

A single invocation of a task, $\tau_i$, is referred to as a job, $j^i_k$. Each job has a remaining execution time, $c^i_k$, a release time, $r^i_k$, and an absolute deadline, $d^i_k$. For periodic tasks, a job is released at every period, $T_i$, (i.e. $r^i_{k+1} = r^i_k + T_i$) with an absolute deadline of $d^i_k = r^i_k + T_i$. Each occurrence of an aperiodic task is also a job with the mentioned properties.

The remaining execution time of a job is initialized as the WCET, and the value is decreased by $f_{slot}$ every time the job is executed in a slot. The normalized frequency, $f_{slot}$, is the ratio of the frequency of the core during the slot execution relative to the maximum frequency (i.e. $f_{slot} = F_{slot}/F_{max}$).

%% file: Section_Scheduler.tex
\section{Scheduler}
\label{section:scheduler}
We assume a time-triggered schedule has been constructed for each core with information on the release time, deadline, and WCET of the VMs and take this as input for the scheduler. Our method is agnostic to the algorithm used to create the schedule. We then use the same concepts presented in slot shifting \cite{fohler1995joint} to provide flexibility in a time-triggered schedule.

\begin{figure}[ht]
\centering
\vspace{-0.5cm}
\begin{subfigure}[ht]{\linewidth}
\includegraphics{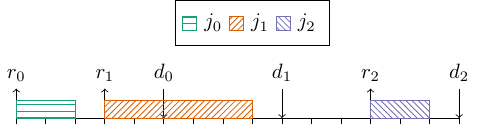}
\caption{Initial scheduling table with WCETs.}\label{sfig:sched_table_initial}
\end{subfigure}
\begin{subfigure}[ht]{\linewidth}
	\hspace*{8pt}\includegraphics[width=0.8\columnwidth]{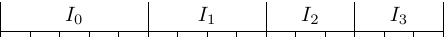}
\caption{Intervals derived from job release times and deadlines.}\label{sfig:sched_table_intervals}
\end{subfigure}
\begin{subfigure}[ht]{\linewidth}
	\hspace*{8pt}\includegraphics[width=0.8\columnwidth]{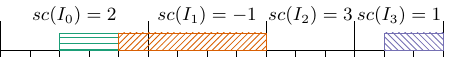}
\caption{Calculation of spare capacity values.} \label{sfig:sched_table_spare_cap}
\end{subfigure}
\vspace{-0.15cm}
\caption{Construction of annotated scheduling table.}

\label{fig:sched_table}
\vspace{-0.25cm}
\end{figure}

A table is created from the input schedule of each core. The table is then annotated with information about the flexibility and leeway available in the schedule. An example of how this is done is shown in Figure \ref{fig:sched_table}.

The entire scheduling table is divided into groups of consecutive slots referred to as a capacity interval ($I$), or simply an interval, as shown in Figure \ref{sfig:sched_table_intervals}. The division is based on the jobs' release times and deadlines. Every unique deadline defines the end of an interval. Each job with the same deadline as the end of an interval belongs to that interval. The start of the interval is the maximum of either the end of the previous interval or the earliest release time of the jobs belonging to the interval. Consecutive slots that do not have a job ready for execution are grouped into a single empty interval.

Each interval has an associated spare capacity value which represents execution time in the interval that is not committed to the jobs in the schedule. This is calculated from the length of the interval, the WCETs of the jobs belonging to the interval, and the spare capacity of the succeeding interval as shown in Equation \ref{eq:sc}. If the spare capacity of an interval is negative, this means that it borrows slots from one or more previous intervals to execute jobs that belong to it. An example of the spare capacity calculation is illustrated in Figure \ref{sfig:sched_table_spare_cap}.
\begin{align}
    sc(I_i) = |I_i| - \sum_{T_j \in I_i} C_j + min(sc(I_{j+1}), 0) \label{eq:sc}
\end{align}

The spare capacity values are regularly updated during runtime, as shown in Algorithm \ref{alg:sc_update}. At every slot, execution time is lost for the current interval, so the spare capacity of the current interval is decremented by 1. If a job, $j$, is executed for a slot, this also frees up execution time for the job's interval, $I_j$. If the job is executed at maximum frequency, a full slot is freed from the job's interval, so it's spare capacity, $sc(I_j)$, is increased by 1. At less than maximum frequency, only part of a slot's execution time is freed --- the same amount that the job's remaining execution time is decreased by, $f_{slot}$. However, this extra execution time can only be used by the job itself (to execute at less than maximum frequency) since only one job is executed at each slot. We refer to this freed partial slot as the \textit{reserved spare capacity} of the job, $reserved\_sc(j)$, and it accumulates every time a job is executed at less than maximum frequency. When the reserved spare capacity of a job reaches 1 or greater, this means a full slot has been freed, which can now be used by other jobs. This is then converted to the spare capacity of the job's interval by incrementing the interval's spare capacity by 1 and decrementing the job's reserved spare capacity by 1. If the interval borrows spare capacity from one or more previous intervals, the spare capacity of the previous intervals that it borrows from is also incremented.

\begin{algorithm}[t]
\caption{Updating spare capacity values}
\label{alg:sc_update}
\begin{algorithmic}
\Function{$ssmod\_timer\_cb(struct hrtimer *timer)$}{}
    \State $sc(I_{curr}) := sc(I_{curr}) - 1$
    \State $reserved\_sc(j) := reserved\_sc(j) + f_{slot}$
    \If{$reserved\_sc(j) \ge 1$}
	\State $reserved\_sc(j) := reserved\_sc(j) - 1$
	\State $prev\_sc := sc(I_j)$
	\State $sc(I_j) := sc(I_j) + 1$
	\State $I_i := I_{curr}$
	\While{$prev\_sc < 0$}
	    \State $I_i := I_{i-1}$
	    \State $prev\_sc := sc(I_i)$
	    \State $sc(I_i) := sc(I_i) + 1$
	\EndWhile
    \EndIf
\EndFunction
\end{algorithmic}
\end{algorithm}

During runtime, the scheduler makes a scheduling decision at every slot. Jobs with earlier deadlines are given higher priority. If there are no jobs to run based on the scheduling table, best-effort/non-critical VMs can be executed.

\subsection{Adding VMs and handling Event-Triggered Tasks}
Slot shifting has a method for integrating new jobs into the scheduling table called the guarantee algorithm. Regenerating new scheduling tables for when a new VM needs to be added can be a relatively long process. To quickly execute new VMs, the guarantee algorithm is used to integrate the jobs of a new VM to the existing schedule.

There are also applications whose executions are unpredictable and intermittent, where it would be more suitable to schedule them on demand rather than regularly. These can be modelled as aperiodic jobs and integrated into the schedule as needed with the acceptance test and guarantee algorithm.

The guarantee algorithm first conducts an acceptance test, which checks if the job can be added to the schedule and be completed within its deadline. This is done by tallying up the spare capacities up to the deadline of the job. If there is enough spare capacity for the worst-case execution time of the job, then it can be accepted. If there isn't enough spare capacity, the scheduler attempts to delegate the job to another core using the negotiation-based acceptance test global algorithm described in \cite{schorr2015adaptive}.

After a job is accepted, the new job is added to the scheduling table. A new interval is created (if needed) and spare capacity values are updated. This guarantees that the new job's worst-case execution time is considered as other jobs are added or the energy-aware techniques described in the next sections use up the spare capacity.

\subsection{Energy-aware slot shifting extension}

The following sections outline our extension to the slot shifting algorithm to use energy-reduction techniques. We describe two separate algorithms, one using DPM (EASS-DPM) and another using DVFS (EASS-DVFS).

\subsubsection{EASS-DPM}
This algorithm attempts to combine idle slots to maximize energy savings by going to a deeper sleep state and minimizing the number of sleep transitions.

The algorithm executes all jobs that are able to execute until completion, and then postpones the execution of the upcoming scheduled jobs as much as possible. When there are no jobs ready to execute in the upcoming slot, the idle period duration is calculated based on the spare capacity of the current interval and possibly succeeding intervals. The algorithm for calculating the idle duration is show in Algorithm \ref{alg:lp_duration}.
\begin{algorithm}[t]
\caption{Low Power Duration}
\label{alg:lp_duration}
\begin{algorithmic}
\Function{$low\_power\_duration$}{}
    \State $duration := sc(I_{curr})$
    \If{$rem\_exec(I_{curr}) == 0$}
        \State $I_i := I_{curr+1}$
        \State $duration := duration + max(0, sc(I_i))$
        \While{$empty(I_i)$}
            \State $I_i := I_{i+1}$
            \State $duration := duration + max(0, sc(I_i))$
        \EndWhile
    \EndIf
    \State \Return $duration$
\EndFunction
\end{algorithmic}
\end{algorithm}

At minimum, the core can be idle for the duration of the spare capacity of the current interval, $I_{curr}$. If there are no jobs in the current interval that have remaining execution times, the core can also be idle for the spare capacities of subsequent intervals up to and including the first interval that has a job that has remaining execution time. This results in the job with the next closest deadline to be executed as late as possible with enough slots for its WCET.

\subsubsection{EASS-DVFS}
This algorithm attempts to use the available spare capacity to execute the current task at the lowest frequency without violating any deadlines.

At each slot, the frequency, $f_{slot}$, to run the core with is calculated. If there is no job to execute in the upcoming slot, the core frequency is set to minimum. If there is a job, $j_k$, to execute, the available spare capacity that the job can use is calculated as shown in Algorithm \ref{alg:avail_spare_cap}. This consists of the job's reserved spare capacity, the spare capacity of the job's interval, and, if the job's interval is not the current interval, the spare capacities of the current and subsequent intervals that have spare capacity and no longer have jobs to execute.
\begin{algorithm}[t]
\caption{Available Spare Capacity}
\label{alg:avail_spare_cap}
\begin{algorithmic}
\Function{$available\_spare\_capacity$}{}
	\State $spare\_capacity := max(0, sc(I_j)) + reserved\_sc(j_k)$
	\State $I_i := I_{curr}$
	\While{$I_i != I_j \land rem\_exec(I_i) == 0 \land sc(I_i) > 0$}
		\State $spare\_capacity := spare\_capacity + sc(I_i)$
		\State $I_i := I_{i+1}$
	\EndWhile
	\State \Return $spare\_capacity$
\EndFunction
\end{algorithmic}
\end{algorithm}

With the available spare capacity, we then calculate the frequency, $f_{slot}$, that the job, $j_k$, will be executed at using Equation \ref{eq:freq}. This spreads the remaining execution time, $c_k$, of the job to use up the available spare capacity.
\begin{align}
	f_{slot} = \frac{c_k}{c_k + \text{Available Spare Capacity}} \label{eq:freq}
\end{align}

%% file: Section_Linux_Implementation.tex
\section{Linux Implementation}
\label{section:linux_implementation}
Our work builds upon TT scheduling class for the Linux kernel proposed in \cite{gala2023joint}.
We implemented our scheduler extensions on the Linux kernel (v5.19.9) with the PREEMPT\_RT patch.
The available CPU cores are divided into managed cores and non-managed cores. Processes that are managed by the scheduler are executed on the managed cores, while other processes may only run on the non-managed cores. To enforce this, the kernel command line parameters \emph{isolcpus} and \emph{irqaffinity} were used.

\begin{figure}[ht]
\centering
\vspace{-0.3cm}
\includegraphics[width=0.8\columnwidth]{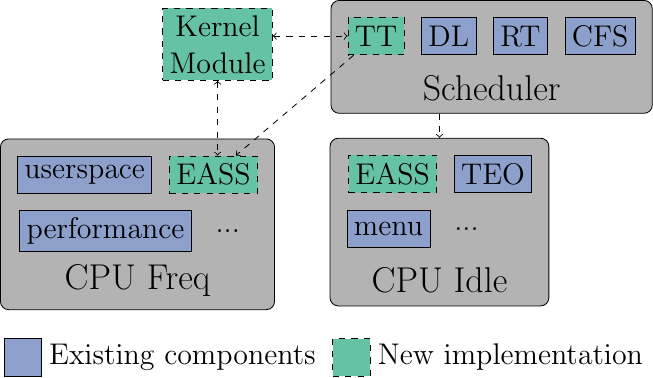}

\caption{Linux implementation overview.}
\vspace{-0.3cm}
\label{fig:implementation_arch}
\end{figure}

The implementation for the energy-aware scheduler consists of the following parts: a Linux scheduling class, a Linux kernel module, and CPUFreq and CPUIdle governors. An overview of the implementation and how it interacts with the core Linux kernel is shown in Figure \ref{fig:implementation_arch}.

Scheduling decisions in the Linux kernel are handled by scheduling classes. Within the core Linux scheduler's \_\_schedule() function, the scheduling classes are polled one by one, by order of priority, if they have a process to schedule. If higher priority scheduling classes return a process to be scheduled, this will be scheduled over processes belonging to a lower priority scheduling class. A new scheduling class was implemented for the energy-aware scheduler and was given the highest priority. This scheduling class performs two main functions: triggering the rescheduling of a core at the start of each slot, and returning the correct process to be scheduled in the upcoming slot.

Most of the scheduling algorithm logic is implemented in the kernel module so that the scheduling class can be used with other scheduling algorithms without needing to recompile the kernel. The kernel module also implements a sysfs interface that lets users control the scheduler, as well as a debugfs interface for collecting logs and statistics.

The CPUIdle subsystem of the Linux kernel handles switching to low power states. When a core goes idle, the subsystem is invoked to select and switch to the appropriate low power state. The CPUIdle subsystem has CPUIdle governors which handles the logic for selecting the low power state. A new CPUIdle governor was created which is used for the EASS-DPM scheduler.

The CPUFreq subsystem of the Linux kernel handles setting the core frequencies of the CPU. Like the CPUIdle subsystem, it also has CPUFreq governors which handles the logic for the frequency switching. A new CPUFreq governor was created to be used for the EASS-DVFS scheduler. Its function is to translate the calculated normalized frequency to a frequency that's available on the platform, as well as interface with the CPUFreq subsystem to change the core frequency.

\subsection{Runtime Operation}
\begin{figure}[ht]
\centering
\vspace{-0.3cm}
\includegraphics[width=0.8\columnwidth]{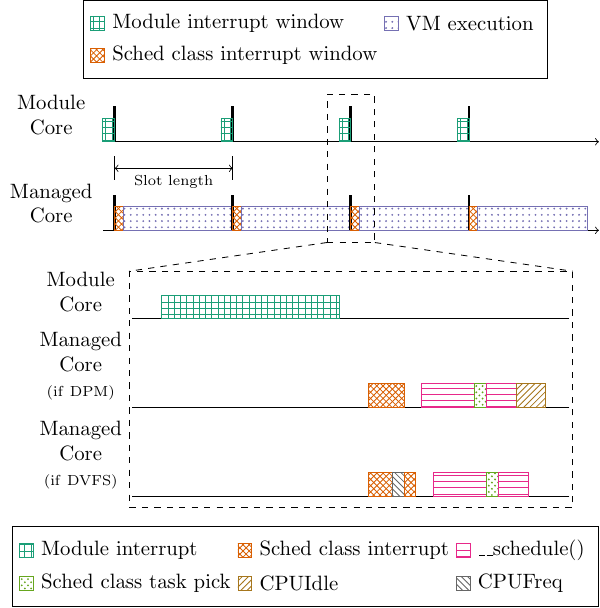}

\caption{Interrupt execution on each slot.}
\vspace{-0.3cm}
\label{fig:slot_interrupts}
\end{figure}

The runtime operation of the scheduler is driven by the hrtimer interrupts of the scheduling class and the kernel module. These interrupts execute relative to the start of every slot, as shown in Figure \ref{fig:slot_interrupts}.

The kernel module sets up an hrtimer interrupt on a non-managed core, referred to as the module core, which runs before the start of every slot. This goes through the scheduling algorithm for each core. It handles moving jobs to different states, accepting new jobs by running the guarantee algorithm, maintaining the spare capacity values of the intervals, and selecting which job to run on each core which it then passes to the scheduling class. With EASS-DPM, if the core will be idle in the upcoming slot, the idle period duration is calculated and passed to the scheduling class. For EASS-DVFS, the ideal normalized frequency is calculated by the module. This value is passed to the CPUFreq Governor, which finds the lowest available frequency that's higher than or equal to the calculated frequency and stores it for the upcoming slot. 

The scheduling class has hrtimer interrupts on each managed core, which fires at the start of each slot. The interrupt checks the scheduling decision made by the kernel module and marks the core for rescheduling if necessary. This, then, starts the scheduling procedure of the Linux core scheduler. The interrupt also tells the CPUFreq governor to change the core frequency in the EASS-DVFS version. With EASS-DPM and if the core is set to be idle for longer than two slots, the next interrupt for that core is delayed until the end of the idle period. Since all other processes and interrupts are configured to run on the housekeeping core, the CPUIdle governor can use the functions made available by the CPUIdle subsystem to determine the time until the next timer interrupt and select the most appropriate low power state. There is no direct communication between the CPUIdle governor and the rest of the EASS-DPM scheduler.

%% file: Section_Evaluation.tex
\section{Evaluation}
\label{section:evaluation}
The scheduler was evaluated on a Dell R640 server with an Intel Xeon Gold 5218 processor. The processor has 16 cores with a max turbo frequency of 3.9 GHz, but this was limited to the base frequency of 2.3 GHz. Each core's frequency can be set independently within a range of 1.0 GHz and 2.3 GHz, with increments of 100 MHz. For our tests, CPU0 was set as a non-managed core while the other cores, CPU1 to CPU15, were managed by the scheduler.

We used the UUnifast algorithm \cite{bini2005measuring} to generate the VM specifications for the experiments. We specify parameters for the algorithm such as the total target utilization, WCET range, and period range. Its output is a set of files containing the properties of the jobs of the VMs. When running the experiment, this output is parsed and fed to the kernel module's sysfs interface.

The performance of the scheduler in terms of power consumption and scheduler overhead was evaluated. We tested three versions of the scheduler: a base slot shifting scheduler without any energy-saving techniques (BSS), the EASS-DPM scheduler, and the EASS-DVFS. For BSS and EASS-DVFS, a CPUIdle governor that doesn't put the cores in a low power state was used. The userspace CPUFreq governor was used for BSS and EASS-DPM, and all cores were set to run at maximum frequency.

\subsection{Power consumption}
To evaluate performance in terms of energy-savings, we generated several test cases with the parameters shown in Table \ref{tab:energy_tset_params}. VMs that are already integrated into the schedule have a total utilization ranging from 20\% to 80\% with increments of 10\%. New jobs that need to be added to the schedule during runtime have a total utilization of 10\%, 20\%, or 50\%. Ten test cases were generated for each combination of the utilization of the scheduled VMs and the utilization of new jobs. Each schedule was 1800 to 2200 slots long and was tested with 3 different slot lengths: 1ms, 3ms, and 10ms. Each test case was run 3 times for each slot length.

\begin{table}[ht]
\centering\vspace{-0.2cm}
\begin{tabular}{ |c|c|c| }
	\hline
	Parameter & Scheduled VMs& New Jobs\\
	\hline
	WCET & [1,15] & [10,15] \\ 
	Period & [15,50] & [10,15]\\
	Utilization & [20 - 80]\% & [10,20,50]\% \\
	\hline
\end{tabular}
\vspace{-0.1cm}
\caption{Parameters of test cases in the energy-usage test.}
\label{tab:energy_tset_params}\vspace{-0.2cm}
\end{table}

The CPU's energy consumption was measured using the Linux kernel's power capping framework \cite{pcapframework}. The \textit{energy\_uj} attribute of the CPU package was recorded at the start, and end of every schedule execution and the difference was divided by the duration of the schedule to get the power measurement.
\begin{figure}[ht]
\centering\vspace{-0.4cm}
\begin{subfigure}{\linewidth}
\centering
\includegraphics[width=0.8\columnwidth]{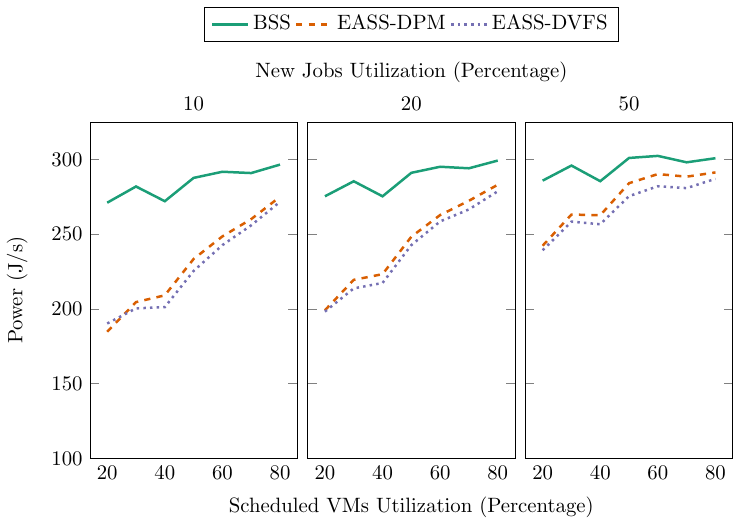}\vspace{-0.1cm}
\caption{Average power consumption.}

\label{sfig:power_measurements}
\end{subfigure}
\begin{subfigure}{\linewidth}
\centering
\includegraphics[width=0.8\columnwidth]{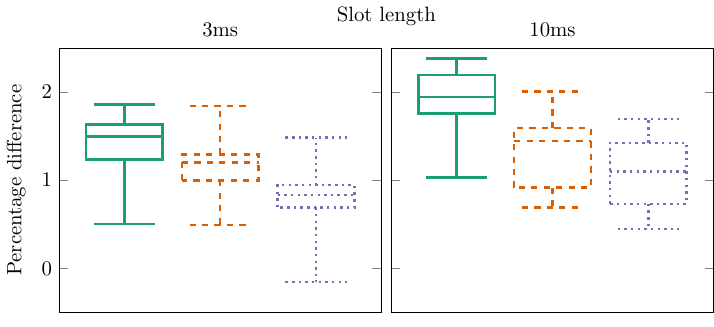}
\vspace{-0.1cm}
\caption{Power difference compared to 1ms slot length.}
\label{sfig:power_slotlen_diff}
\end{subfigure}
\vspace{-0.1cm}
\caption{Power consumption results.}

\label{fig:power_results}\vspace{-0.3cm}
\end{figure}

Figure \ref{sfig:power_slotlen_diff} shows the power percentage difference when using a slot length of 3ms and 10ms, with the difference in average power of each utilization combination as a data point. There is a decrease in power consumption as the slot length gets larger, but the difference is minute. This difference could be due to more frequent interrupts and context switches.

Figure \ref{sfig:power_measurements} shows the average power consumption for each utilization combination. There is a significant decrease in power consumption when comparing the energy-aware schedulers to that of the base scheduler, with as much as a 31.85\% and 29.81\% decrease on the lowest utilization values for EASS-DPM and EASS-DVFS respectively. As can be expected, the difference decreases as utilization increases. At 50\% total utilization, there was a decrease of roughly 23\% for EASS-DPM and 28\% for EASS-DVFS. This tapers down to a 3.16\% difference at the highest utilization level for EASS-DPM and a 4.58\% difference for EASS-DVFS.

\subsection{Scheduling overhead}

To measure the scheduler overhead, 50 test cases were generated with the parameters shown in Table \ref{tab:ohead_exp_tset_params}. The schedule durations for each test case ranged from 480 to 520 slots, with a slot length of 3ms. Each test case was run 5 times for each scheduler.

\begin{table}[ht]
\centering
\begin{tabular}{ |c|c|c| } 
	\hline
	Parameter & Scheduled VMs & New Jobs\\
	\hline
	WCET & [1,15] & [10,15] \\ 
	Period & [15,50] & [10,15]\\
	Utilization & 50\% & 50\% \\
	\hline
\end{tabular}\vspace{-0.1cm}
\caption{Parameters of the test cases in the overhead test.}
\label{tab:ohead_exp_tset_params}\vspace{-0.4cm}
\end{table}

\begin{figure}[ht]
\centering
\begin{subfigure}{\linewidth}
	\centering
	\includegraphics[width=0.8\columnwidth]{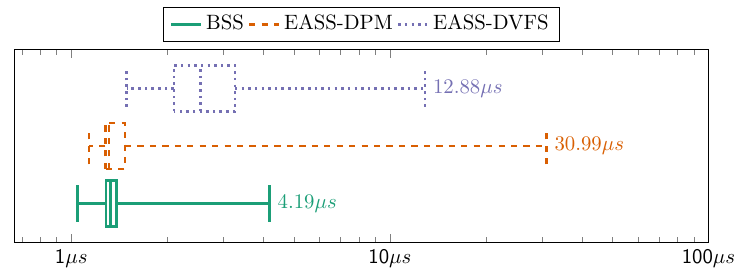}
	\vspace{-0.1cm}

	\caption{Switch duration.}

	\label{sfig:ohead_mcores_sched_duration}
\end{subfigure}
\begin{subfigure}{\linewidth}
	\centering
	\includegraphics[width=0.8\columnwidth]{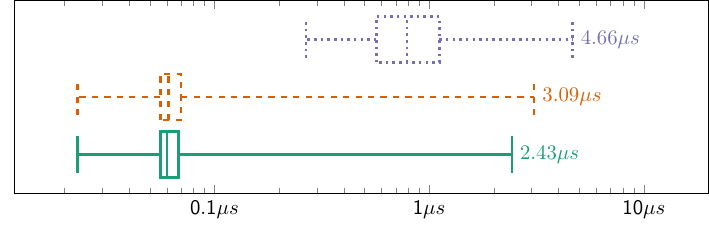}
	\vspace{-0.1cm}
	\caption{Scheduling class interrupt.}

	\label{sfig:ohead_mcores_tick}
\end{subfigure}
\begin{subfigure}{\linewidth}
	\centering
	\includegraphics[width=0.8\columnwidth]{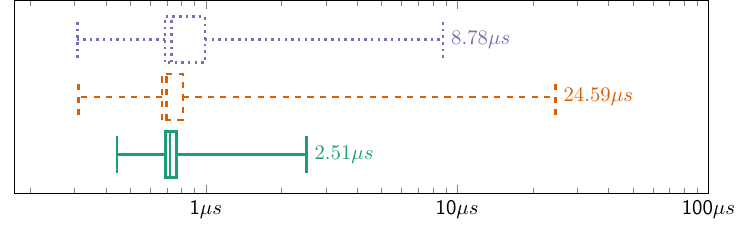}
	\vspace{-0.1cm}
	\caption{\_\_schedule() function.}

	\label{sfig:ohead_mcores_sched}
\end{subfigure}
\vspace{-0.3cm}
\caption{Overhead measurements for managed cores.}
\label{fig:ohead_mcores}
\end{figure}

Figure \ref{sfig:ohead_mcores_sched_duration} shows the measurements from the start of the scheduling class interrupt up until the end of the \_\_schedule() function. We see that for the majority of cases (i.e. within the 1st and 3rd quartile), EASS-DVFS has a higher switch duration when compared to BSS and EASS-DPM. On average, EASS-DVFS had a switch duration of 2.74$\mu s$ as compared to the 1.44$\mu s$ and 1.52$\mu s$ of BSS and EASS-DVFS respectively. One reason for this is that EASS-DVFS switches the core frequency during the scheduling class interrupt, which results in more overhead, as can be seen in Figure \ref{sfig:ohead_mcores_tick}. Another reason, is that EASS-DVFS lowers the frequency of the managed cores, which also slows down the scheduler.

\begin{figure}[ht]
\centering
\vspace{-0.4cm}
\begin{subfigure}{\linewidth}
	\centering
	\includegraphics[width=0.7\columnwidth]{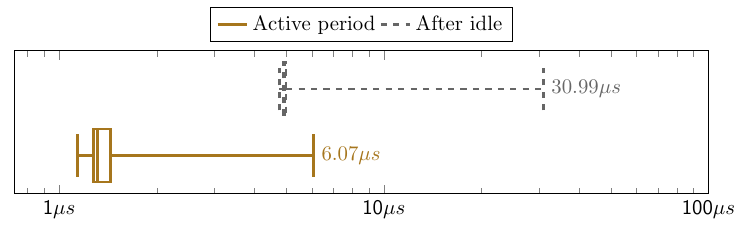}
	\vspace{-0.1cm}
	\caption{Switch duration.}
	\label{sfig:ohead_mcores_dpm_sched_duration}
\end{subfigure}
\begin{subfigure}{\linewidth}
	\centering
	\includegraphics[width=0.8\columnwidth]{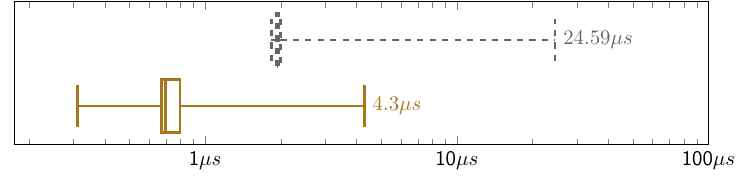}
	\vspace{-0.1cm}
	\caption{\_\_schedule() function.}

	\label{sfig:ohead_mcores_dpm_sched}
\end{subfigure}
\vspace{-0.3cm}
\caption{Overheads for EASS-DPM on managed cores.}
\label{fig:ohead_mcores_dpm}\vspace{-0.2cm}
\end{figure}

It is also apparent in Figure \ref{fig:ohead_mcores} that, even though EASS-DPM had comparable values to the base scheduler in the average case, it had significantly higher values above the 3rd quartile. We've observed that these high values occur right after returning from a low power state. This can be seen in Figure \ref{fig:ohead_mcores_dpm}, which shows the overhead data for EASS-DPM but with the instances immediately after returning from a low power state separated from the rest of the data. Even with the highest observed value of 30.99$\mu s$, this is still only about 1\% of the slot length in this test.

\begin{figure}[ht]
\centering
\begin{subfigure}{\linewidth}
	\centering
	\includegraphics[width=0.8\columnwidth]{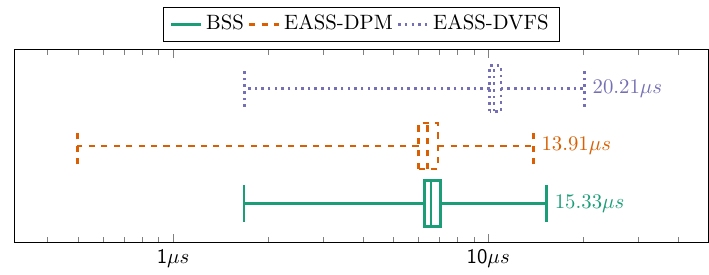}
	\vspace{-0.1cm}
	\caption{Overall module interrupt duration.}

	\label{sfig:ohead_mod_interrupt}
\end{subfigure}
\begin{subfigure}{\linewidth}
	\centering
	\includegraphics[width=0.8\columnwidth]{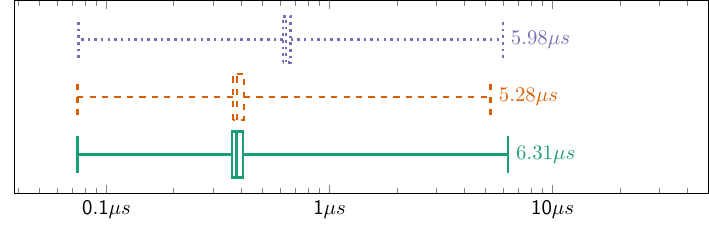}
	\vspace{-0.1cm}
	\caption{Per core duration.}

	\label{sfig:ohead_mod_per_core}
\end{subfigure}
\vspace{-0.3cm}
\caption{Overhead measurements on the module core.}
\label{fig:ohead_mod}
\end{figure}

In Figure \ref{sfig:ohead_mod_interrupt} we compare the module interrupt durations between the three versions of the scheduler. Note that the module interrupt runs on a non-managed core and, therefore, does not affect any of VMs managed by the scheduler.

The values from EASS-DPM are comparable to the base scheduler in the average case, and has significantly lower values below the 1st quartile. This is because the scheduler algorithm is not run on cores that are in a low power state. As shown in the per core measurements in Figure \ref{sfig:ohead_mod_per_core}, EASS-DPM is quite close to the base scheduler on a per core basis.

\begin{figure}[t]
\centering
\vspace{-0.2cm}
\includegraphics[width=0.8\columnwidth]{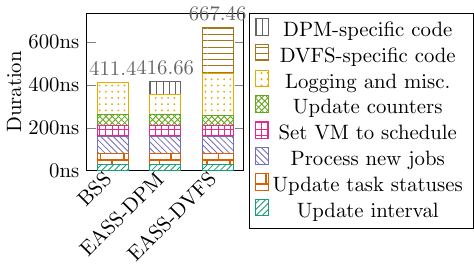}
\vspace{-0.2cm}
\caption{Breakdown of average per core measurements.}

\label{fig:ohead_mod_breakdown}\vspace{-0.1cm}
\end{figure}

In contrast, significantly higher values were measured for the EASS-DVFS. The added functionality of the DVFS algorithm amounts to a significant increase in the overhead measurements for the scheduler module, as can be seen in Figure \ref{fig:ohead_mod_breakdown}, which shows a breakdown of the overhead values for different parts of the scheduler algorithm. This extra overhead for the EASS-DVFS includes calculating for the ideal frequency, as well as sending it to the CPUFreq governor to find a suitable frequency that's available on the CPU.

%% file: Section_Conclusion.tex
\section{Conclusion}
\label{section:conclusion}

We presented an energy-aware time-triggered scheduler for cloud nodes based on Kernel Virtual Machine (KVM) in conjunction with Linux. The scheduler uses concepts from the slot shifting algorithm to provide flexibility at run-time for adding real-time Virtual Machines (VMs) to the offline scheduling table without requiring a high overhead table regeneration process. In addition, the scheduler also uses the slack in the schedule to apply energy-saving techniques without compromising the requirements of the real-time VMs.

We presented two versions of the scheduler as an extension to slot shifting: EASS-DPM and EASS-DVFS. We implemented both versions on Linux for use with KVM.
We performed experiments on server-grade hardware with Intel Xeon $2^{nd}$ generation processors (Cascade Lake) to evaluate both the scheduler implementations for energy savings and scheduling overheads.

The evaluation shows a significant reduction in power consumption from the two energy-aware algorithms presented, with as much as a 31.85\% difference in the lowest utilization level and around 23\%-28\% at 50\% utilization. Both versions performed similarly in this aspect, with EASS-DVFS performing slightly better in most cases.
In terms of overhead, we observed higher values with EASS-DVFS in the average case, but EASS-DPM had higher values in extreme cases. However, even in the most extreme case, we observed the overhead to be only 1\% of the slot length.

In this paper, we assumed VMs always execute until their worst-case execution time. However, this is usually not the case in actual practice. In the future, we will account for the actual execution time of VMs to save energy further. Moreover, we plan to combine the DPM and DVFS scheduler versions to achieve even better performance in terms of energy. 